\documentclass[amssymb,aps,twocolumn,showpacs, superscriptaddress, nofootinbib]{revtex4}
\usepackage[]{graphicx}
\usepackage[]{amsmath}
\usepackage{hyperref}
\usepackage{subfigure}

\def\ket#1{| #1 \rangle}

\def\rme{\mathrm{e}}

\def\cH{\mathcal{H}}

\def\cP{\mathcal{P}}

\def\cS{\mathcal{S}}

\def\bbZ{\mathbb{Z}}

\def\eq#1{Eq.~\eqref{eq:#1}}
\def\fig#1{Fig.~\ref{fig:#1}}
\def\be{\begin{equation}}
\def\ee{\end{equation}}
\def\BE{\begin{equation}}
\def\EE{\end{equation}}
\newcommand\BF{\begin{figure}}
\newcommand\EF[2]{\caption{#1}\label{#2}\end{figure}}

\begin{document}

\title{Kitaev's $\mathbb{Z}_d$-Codes Threshold Estimates}
\author{Guillaume Duclos-Cianci}
\affiliation{D\'epartement de Physique, Universit\'e de Sherbrooke, Sherbrooke, Qu\'ebec, J1K 2R1, Canada}
\author{David Poulin}
\affiliation{D\'epartement de Physique, Universit\'e de Sherbrooke, Sherbrooke, Qu\'ebec, J1K 2R1, Canada}

\date{\today}

\begin{abstract}
We study the quantum error correction threshold of Kitaev's toric code over the group $\bbZ_d$ subject to a generalized bit-flip noise. This problem requires novel decoding techniques, and for this purpose we generalize the renormalization group method we previously introduced in \cite{DP09a,DP10a} for $\bbZ_2$ topological codes. 
\end{abstract}


\maketitle

Kitaev's topological code (KTC) \cite{K03a} on qubits is the archetypical topological code and has been extensively studied. As explained in Kitaev's original paper \cite{K03a}, this construction applies to any group. Much less is known about these generalizations, and in this paper we investigate the quantum error correction (QEC) thresholds of the KTCs built with the groups $\bbZ_d$, where $d\geq2$. We label these as $\bbZ_d$-KTC, so the original code on qubits corresponds to $\bbZ_2$-KTC. 

As explained in \cite{DKLP02a}, $\bbZ_2$-KTC can be decoded by a binary perfect matching algorithm \cite{E65a}, since every particle is its own anti-particle in this model. Because this is not the case for $d > 2$, other techniques are required and for this purpose we generalize the renormalization group (RG) soft decoder that we introduced in \cite{DP09a,DP10a}. Our numerical simulations show that the threshold increases monotonically with $d$ and  appears to follow the general trend of the qudit hashing bound.

This paper is organized as follows. First, we introduce a generalized Pauli group (see \cite{K96a,G99a} for more details), stabilizer codes, and $\bbZ_d$-Kitaev's toric code. Next, we briefly review the decoding problem of these systems and show how the RG decoder applies in this case. Finally, we present the numerical results and close with a discussion.

\section{$\mathbb{Z}_d$ generalization of Kitaev's toric code}

In this section, we review the definition of $\bbZ_d$-KTC and show that many features of KTC on qubits extend to them. Since we will be working with qudits, we introduce a generalized Pauli group. The Hilbert space of a qudit, $\cH_d$, is spanned by the states $\{\ket{0},\ket{1},\dots,\ket{d-1}\}$. We define the operators $X$ and $Z$ such that
\begin{align}
	X\ket{g} &= \ket{g\oplus1}, \label{eq:defXZ}\\
	Z\ket{g} &= \omega^g\ket{g}, \nonumber
\end{align}
where $0\leq g<d$, ``$\oplus$" denotes addition modulo $d$, and $\omega=\rme^{i 2\pi/d}$. The generalized Pauli group is generated by $X$, $Z$, and a phase, i.e., $\cP_d=\langle \omega,X,Z\rangle$ if $d$ is odd and $\cP_d=\langle \omega^{1/2},X,Z\rangle$ if $d$ is even ($XZ$ has order $2d$ in this case). From the definitions of \eq{defXZ}, we deduce the following properties
\begin{align}
	X^a\ket{g} &= \ket{g\oplus a},\nonumber\\
	Z^a\ket{g} &= \omega^{ag}\ket{g}, \label{eq:properties}\\
	ZX\ket{g} &= \omega XZ\ket{g},\nonumber\\
	Z^aX^b\ket{g} &= \omega^{ab} X^bZ^a\ket{g}.\nonumber
\end{align}
Lastly, we define the $n$-qudit Pauli group $\cP_d^n \equiv \cP_d^{\otimes n}$ as the $n$-fold tensor product of $\cP_d$.

The stabilizer group $\cS$ is an ablian subgroup of $\cP_d^n$. The code is defined as the simultaneous +1 eigenspace of all stabilizers. Note that even though the generalized Pauli operators are unitary, they are not hermitian in general so do not correspond to physical observables. However, the operator $\frac12 (s+s^\dagger)$ is hermitian and can be measured. Since $s$ has eigenvalues $\omega^a$, $\frac12 (s+s^\dagger)$ has eigenvalues $\frac12 (\omega^a+\omega^{-a})=\cos(2\pi a/d)$ which are in one-to-one correspondence with the eigenvalues of $s$.

\BF
	\includegraphics[width=6cm]{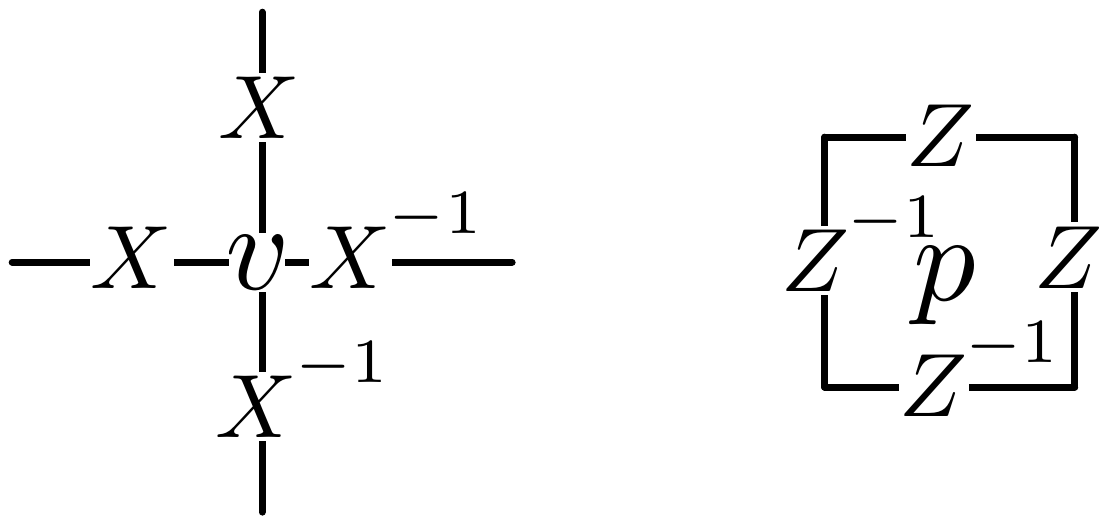}
\EF{$\bbZ_d$-KTC stabilizer generators. To each vertex $v$, we associate an operator $A_v$(left) and to each plaquette $p$, we associate an operator $B_p$ (right).}{fig:stab_gen}

With these definitions  in place, we present a generalization of KTC on qudits, which we call $\bbZ_d$-KTC, using Kitaev's original construction \cite{K03a} on the cyclic groups $\bbZ_d$ with $d\geq2$. The system is a square lattice of linear size $L$ with periodic boundary conditions. Each edge is occupied by a qudit, so there are in total $n = 2L^2$ qudits. We define vertex operators $A_v$ and plaquette operators $B_p$ as shown in \fig{stab_gen}. There is one such operator for each vertex and each plaquette. We verify that they commute using the last line of \eq{properties}. These operators generate the stabilizer group $\cS=\langle A_v,B_p\rangle$ and the code is spanned by the simultaneous +1 eigenstates of the stabilizer generators.

\BF
	\includegraphics[width=6cm]{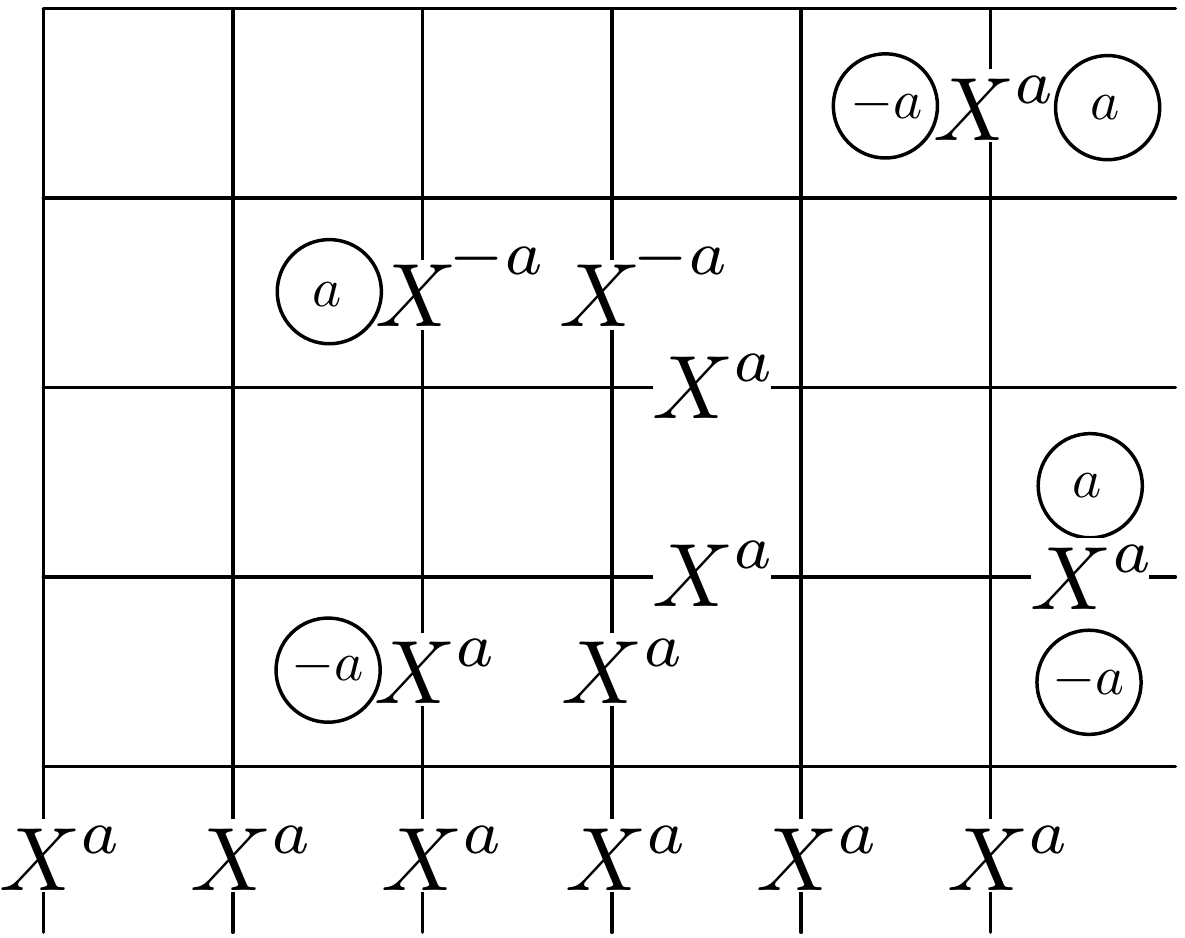}
\EF{Plaquette defects created by the application of some power of $X$. The values $a$ $(-a)$ in the plaquettes are such that the eigenvalue of the corresponding $B_p$ is $\omega^a$ $(\omega^{-a})$. By choosing appropriately the powers of X, we can build string operators with defects only on their endpoints. Non-trivial cocyles of $X^a$ correspond to $\overline{X}^a$ logical operators.}{fig:charges}

Figure \ref{fig:charges} illustrates how applying some power of $X$ on a codestate creates defects on the lattice. Indeed, $X^a$ applied on some qudit does not commute with the two plaquette operators involving that qudit. The eigenvalues of the plaquettes to the north or east of the error will change from 1 to $\omega^a$, and those of the plaquettes to the south or west will change from 1 to $\omega^{-a}$. One can show that the defects thus created are topological charges; we associate the charge $a$ to a plaquette defect corresponding to an eigenvalue $\omega^a$ of that plaquette. With this choice of labeling, the charge group restricted to plaquettes is $\bbZ_d$ with addition.

From these simple facts, it follows that string operators can be built with defects attached only to their endpoints (these strings actually live on the dual lattice, just like in KTC). This requires a careful choice of the powers of $X$ on the qudits along the string such that the total charge in each plaquette is 0 except on its endpoints. For instance, one can adopt the convention that power $a$ is used when heading north or east, and $-a$ when heading south or west.  Moreover, we can verify that non-trivial cocycles (loops on the dual lattice, see \fig{charges}) of any power of $X$ obeying this convention commute with the stabilizer. These operators are not in the stabilizer as all the vertex generators of \fig{stab_gen} are trivial cocyles. It follows that such operators, e.g. the one found at the bottom of \fig{charges}, are logical operators (for any value of $a$).  

A similar analysis holds for defects created by powers of $Z$ operators. In this case, the defects live on vertices  and string operators, on the direct lattice. Also, non-trivial cycles of any power of $Z$ are logical operators. From the form of the logical operators, we directly deduce that there are two qudits encoded in the code space. Again, this is analogous to the case of KTC.

\section{$\bbZ_d$-KTC decoding}

We are now interested in the problem of error correcting $\bbZ_d$-KTCs for $d>2$. In our study, we consider a simple noise model that generalizes the independent symmetric bit-flip channel to qudits\footnote{This noise model can also be seen as emerging from a qudit depolarization channel that maps $\rho \rightarrow (1-q)\rho + q\frac{I}{d}$ when $X$ and $Z$ errors are treated independently, and $p_{\rm phys} = q(1-d^{-1})$.}: with probability $1-p_{\rm phys}$, the qudit remains unaffected and with probability $p_{\rm phys}$, we apply at random (uniformely distributed) one of $X,X^2,\dots,X^{d-1}$. Suppose an error $E\in\cP_d^n$ occurs on a code state. It creates defects on the lattice and by measuring the eigenvalues of every $\frac12(A_v+A_v^\dagger)$ and $\frac12(B_p+B_p^\dagger)$ we can learn the position and charge of each defect. The role of the decoder is to bring the system back in the code space by applying a correcting Pauli operator, $C\in\cP_d^n$. However, care must be taken in choosing an appropriate correcting operation. Indeed, if the operator $CE$ resulting from the combination of the error and the recovery is an element of $\cS$, the state is unaffected. However, if $CE$ is a non-trivial logical operator, then the system is returned to the code space but potentially in a different code state, so the information is corrupted.

Any operator $E\in \cP_d^n$ creating the measured configuration of defects is a potential error. However, we classify these operators by their logical effect on the code space: two operators $E_1, E_2$ with the same configuration of defects are equivalent iff $E_2^\dagger E_1$ has a trivial effect on the code, i.e. $E_1\sim E_2$ iff $E_2^\dagger E_1\in \cS$. Note that since $E_1$ and $E_2$ lead to the same defect configuration, $E_2^\dagger E_1$ creates no defect, or equivalently, $E_1$ creates some defects that $E_2^\dagger$ annihilates.

Given a measured defect configuration, the decoder seeks for the best correction among the set of all errors which would lead to this defect configuration. One strategy would be to identify the error from this set that has the largest probability $\cP(E)$, where the probability of an error is specified by the physical noise model, in our case the symmetric bit-flip channel. This turns out not to be optimal however, because some errors have equivalent effects on all code states. Thus, the decoder should instead seek for the most likely equivalence class of errors. The probability of an equivalence class of errors is obtained by summing over the probability of each error within a class. Given these probabilities, the optimal correction consists in applying the adjoint of any representative of the class with maximal probability.

\section{RG decoder generalization to $\bbZ_d$-KTC}

\begin{figure}
  \subfigure[ ]{
  \includegraphics[width=6 cm]{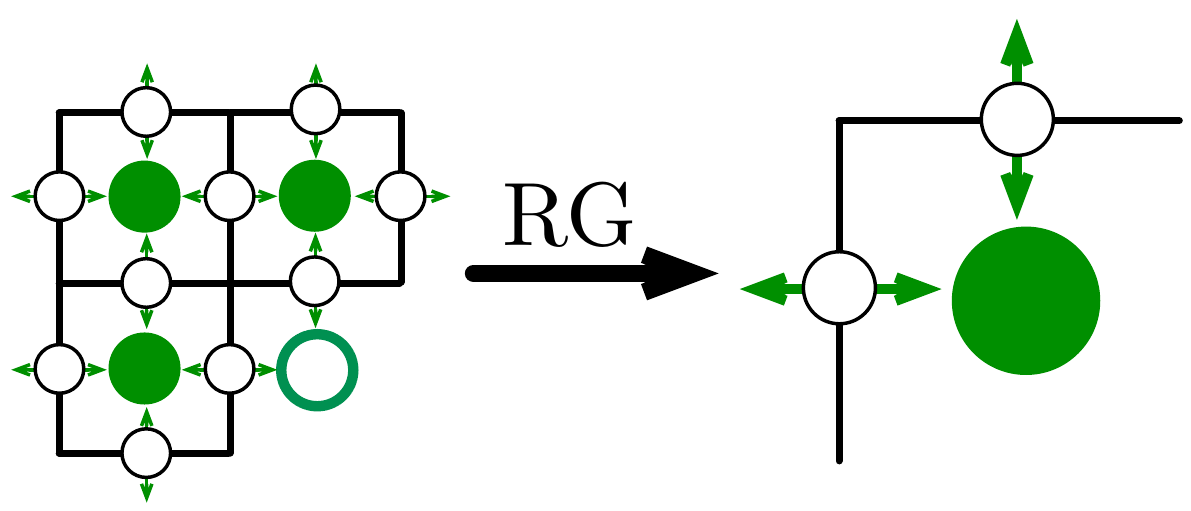}
  \label{fig:UCRGa}}
  \subfigure[ ]{
  \includegraphics[width=2 cm]{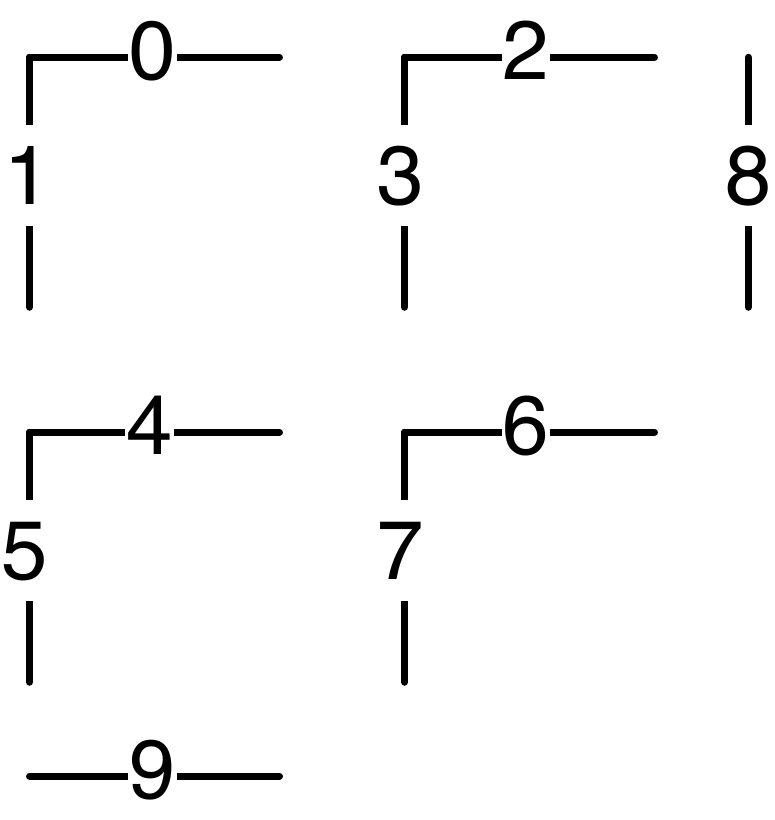}
  \label{fig:UCRGb}
  }
  \caption{(a)  The lattice is cut into unit cells containing ten qudits (edges). The renormalization process takes the defect configuration and the noise model on a unit cell as inputs and outputs a two-qudit distribution (white disks) which corresponds to a probability on the charge flow through the corresponding boundaries. Green disks represent plaquette operators. The plaquette corresponding to the green circle is replaced by the product of all four plaquettes of the unit cell, such that its eigenvalue gives the total charge of the cell. This value is only going to be used in the next round of RG (larger green disk). (b) Labeling convention for qudits in \eq{UCRGbasis}}
  \label{fig:UCRG}
\end{figure}

Unfortunately, the above procedure cannot be realized efficiently in general since the number of errors in each equivalence class scales exponentially with the system size. In \cite{DP09a,DP10a}, we introduced a renormalization group soft decoder (RG decoder) that efficiently approximates the exact calculation (see \cite{BH11a} for a related scheme). The general idea is to cut the lattice into small unit cells (e.g. $2\times2$ sub-lattices) and to ``distill" from each cell an effective two-qubit noise model, c.f. \fig{UCRG}(a). This is realized by keeping track of the flow of charges through the cell and summing over the microscopic details leading to this flow. This has the effect of shrinking the lattice linear size by a constant factor ($k$ for cells of size $k\times k$). Recursing on this process, one can shrink the lattice to a constant, manageable, size where the exact decoding can be performed. With appropriate simple modifications, this method can be used for charges over $\bbZ_d$.

There are two technical difficulties in realizing the above heuristic description, which are both caused by charge conservation. First, because the unit cells share boundaries,  the flow of charge through one boundary of a cell should be equal and opposite to the flow of charge of the corresponding boundary of the neighbouring cell. Thus, the variable corresponding to charge flows in each cell are highly constrained. This problem is easily circumvented by  keeping only  track of the flow of charge through the northern and the western boundary of each cell, i.e. by eliminating this redundancy. 

Second, the sum of the charge flow through the boundaries of a cell must be equal to its total charge, revealed by the syndrome measurement. This once again sets a hard constraint between the variables corresponding to the charge flows, which would in principle require a probability distribution that correlates all the variables of the system. This cannot be realized efficiently, so we  must resort to some approximation. As a first approximation, we choose to ignore the cross-cell correlations, and keep only marginal probabilities on the flows associated to a given cell (we keep a probability distribution that involves the northern and western boundary only). To diminish the effect of these correlations we are neglecting, we let the charge inside a unit cell fluctuate. For each unit cell, we measure all but one of the plaquettes it encloses. This remaining plaquette thus determines the total charge of the unit cell, and indeed we can substitute the corresponding stabilizer generator by a plaquette enclosing the entire unit cell (obtained by multiplying all the plaquette operators contained in the unit cell). This new stabilizer generator represents a renormalized charge. 

This procedure is illustrated on \fig{UCRG}(a) where green disks represent plaquettes that are measured and the  green circle represents the plaquette that is left fluctuating. This green circle is  replaced by the larger, renormalized green disk (on the right) that is used in the next RG step. The white disks on this figure each represent a probability distribution on charge flow, or equivalently a two-qudit probability distribution. 
Thus, after one round of RG, we are left with a smaller lattice and both renormalized charges and renormalized noise models. 

Equation~(\ref{eq:UCRGbasis}) lists a set of generators for all $X$ operators living on a unit cell (see \fig{UCRG}(b) for labelling). This basis will be used to decompose any $X$-type error contained on the unit cell. These operators are defined in accordance to the renormalization process itself as we now explain. The $T_i$ operators are used to build a representative error with the appropriate defect configuration. Indeed, only the $T_i$ operators of \eq{UCRGbasis} do not commute with all three plaquette operators in the unit cell (green disks of \fig{UCRG}(a)). Label the defect configuration on a unit cell as $\vec a=(a_0,a_1,a_2)$, where $a_0$ is the charge of the north-west plaquette, $a_1$ is the charge of the north-east one, and $a_2$ is the charge of the south-west one. Then, the Pauli operator $t(\vec a) = T_0^{a_0}T_1^{a_1}T_2^{a_2}$ creates the defect configuration $\vec a$. Moreover, given a defect configuration $\vec a$, every potential error has to contain this product in its decomposition on basis \eq{UCRGbasis} since only the $T_i$ operators do not commute with plaquettes. The $L_i$ operators characterize the flow of charge through the northen and western boundaries, so the two-qudit ouput distribution of a RG round is precisely the probability distribution over these two operators. The $S_i$ operators are stabilizer operators (or parts of stabilizer generators supported on the unit cell). They only deform strings without changing their defect configuration or their associated charge flow. Lastly, the $E_i$ operators correspond to charge flowing through the southern and eastern boundaries into the plaquette operator that is left out. Thus, they are responsible for the charge fluctuation inside the unit cell and they are summed over.

\begin{align}
	S_0&=X_0X_2^{-1}X_3^{-1}&T_0&=X_4X_7^{-1}\nonumber\\
	S_1&=X_1X_4^{-1}X_5^{-1}&T_1&=X_6\nonumber\\
	S_2&=X_3X_4X_6^{-1}X_7^{-1}&T_2&=X_7^{-1}\nonumber\\
	\label{eq:UCRGbasis}\\
	E_0&=X_6X_8&L_0&=X_2X_6\nonumber\\
	E_1&=X_7^{-1}X_9^{-1}&L_1&=X_5X_7\nonumber
\end{align}

With these definitions, we can formally describe a RG round that starts with a defect configuration $\vec a$, and computes the marginal probability of each $l \in\langle L_0,L_1\rangle$ conditioned on the measured defect configuration,
\begin{align}
	\cP(l)=\sum_{e\in\langle E_0,E_1\rangle}\sum_{s\in\langle S_0,S_1,S_2\rangle} \cP(tles), \label{eq:RG}
\end{align}
where $t=T^{a_0}T^{a_1}T^{a_2}$ is given by the defect configuration and $\cP(tles)$ is the probability assigned to the error $E=tles$ by the noise model. The complexity of decoding a unit cell is given by the number of operators that are considered in \eq{RG}: $|\langle L_0,L_1\rangle|\cdot|\langle E_0,E_1\rangle|\cdot|\langle S_0,S_1,S_2\rangle|$. Since all $L_i$, $E_i$ and $S_i$ have order $d$, the complexity is the constant $d^7$. For different unit cell sizes, the complexity is still a power of $d$, but with a different exponent which depends on the number of qudits in the cell and the number of measured stabilizer generators. Moreover, the number of unit cells to decode in a given round of RG is given by $(L/k)^2$ where $k$ and $L$ are the linear sizes of the unit cell and the global lattice, respectively. Thus, the complexity of a step of RG goes as $d^c(L/k)^2$ for some constants $c$ and $k$ that depend on the choice of unit cell. Of course, the RG calculations on different cells can be executed in parallel.

The procedure we have described above to evade the correlations caused by local charge conservation is only a heuristic, and can be improved using belief propagation (BP). Roughly, the role of BP is to ensure consistency between the marginal probability of qubits located at the boundary of two or more unit cells, e.g. qudits 0, 1, 8 and 9 (see \fig{UCRG}(b) for labeling). First, given a defect configuration inside a unit cell, one can compute the marginal error probability $\cP_q(tles|_q)$ for each qudit $q$, obtained by taking a marginal of $\cP(tles)$. These are called messages and denoted $m^{\rm out}_q(p)$, where $q$ labels a qudit and $p$ is a one-qudit Pauli operator. These outgoing messages are then exchanged between neighbouring cells, and become incoming messages, e.g. a cell $c$ sends to its northern neighbour $c'$ the message $m^{\rm out}_0$ that becomes $m_9^{\rm in}$ in $c'$,  and  receives from $c'$ the message $m^{\rm out}_9$ that becomes $m^{\rm in}_0$ in $c$. Subsequent rounds of messages can be calculated using the received messages, following the prescription
\begin{align}
	m^{\rm out}_q(p)\leftarrow \sum_{l,s,e}\delta(tles|_q,p)\frac{\cP(tles)}{\cP_q(tles|_q)}\prod_{q'\neq q}m^{\rm in}_{q'}(tles|_{q'}),
\end{align}
Here, $q,q'\in\{0,1,8,9\}$, $tles|_q$ is the restriction to qudit $q$ of the Pauli operator $tles$ and $\cP_q$ is the marginal on qudit $q$ of the noise model as above. BP can be iterated a few times (e.g. three rounds) before executing a RG step. This has the effect of replacing \eq{RG} by
\begin{align}
	\cP(l)=\sum_{e\in\langle E_0,E_1\rangle}\sum_{s\in\langle S_0S_1S_2\rangle} \cP(tles)\prod_qm^{\rm in}_q(tles|_q). \label{eq:RGm}
\end{align}

\section{Numerical results} 

\BF
	\includegraphics[width=8cm]{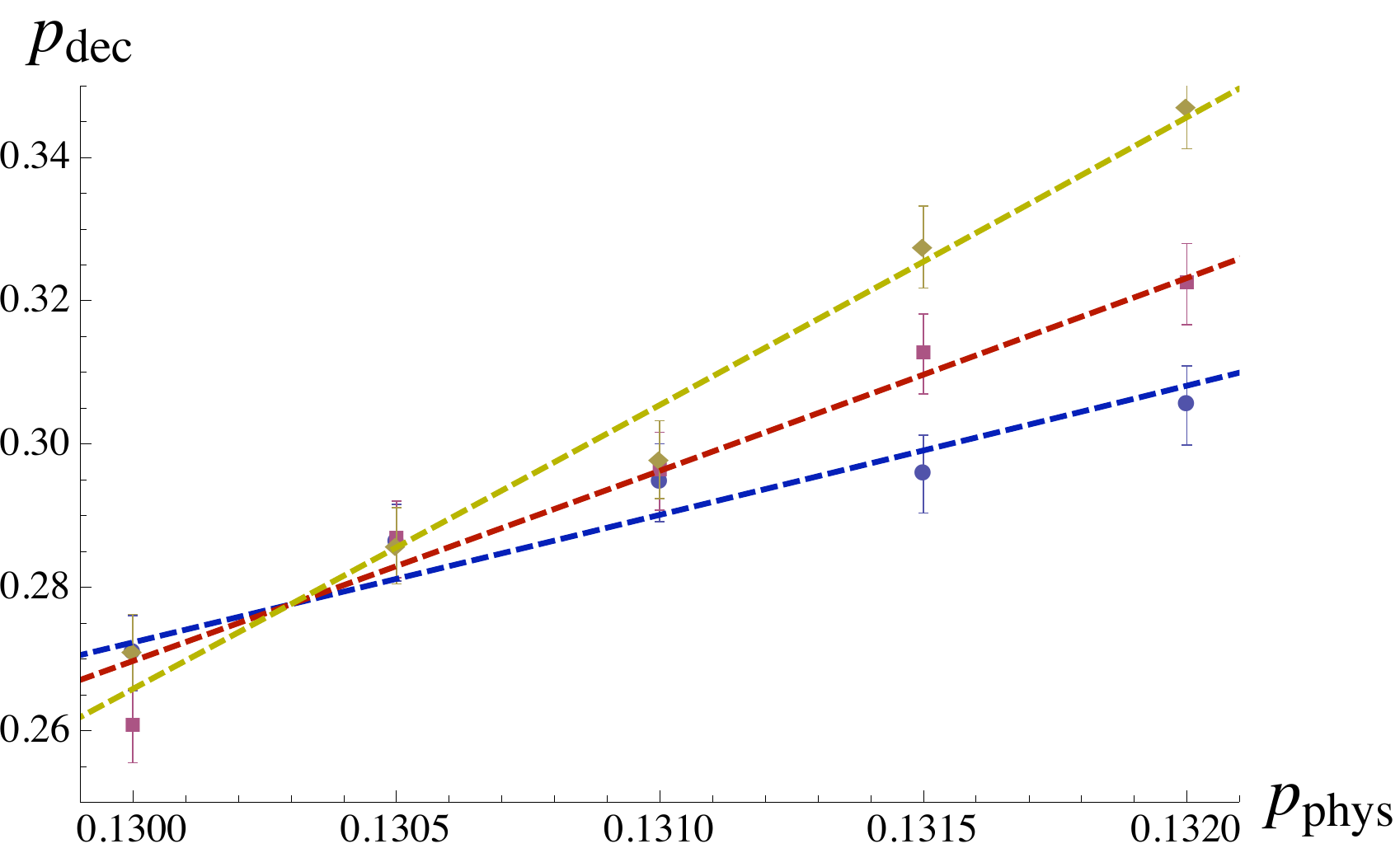}
\EF{Threshold estimation for $\bbZ_3$-KTC. The x-axis represents physical error rate and the y-axis, decoding error rate. The blue dots, red squares and yellow diamonds correspond to $L=32$, $L=64$ and $L=128$ respectively. The fitting curve used is $p_{dec}=(p_{phys}-p_{th})L^{1/\nu}$. In this case, we find $p_{th}=0.13(0)$.}{fig:Z3_pth}

\BF
	\includegraphics[width=8cm]{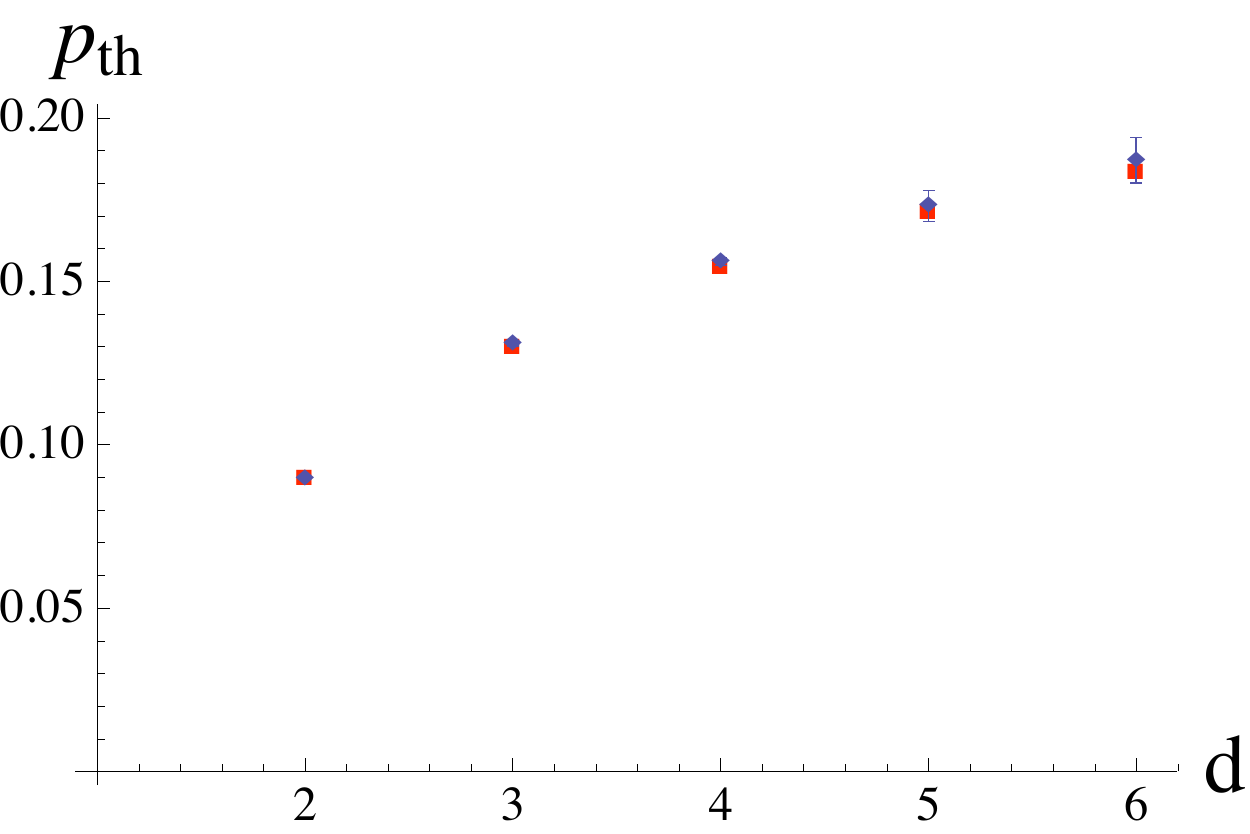}
\EF{The blue diamonds are the values extracted by fitting the threshold values for $2\leq n\leq6$ (see \fig{Z3_pth} for example). The red squares are obtained via the generalized hashing bound (see text) rescaled by a common factor $\alpha = p_{th}(2)/C_2 \approx 0.81$. The error bars are (pessimistically) obtained e.g. by replacing each line in \fig{Z3_pth} by a stripe of width equal to the statistical error bars, and determining the values of $p_{\rm phys}$ above and below the crossing point where the strips cease to overlap. We do not report the fitting parameter $\nu$ because they are too sensitive to statistical fluctuations and therefore unreliable in our study.}{fig:HB_Results}

In this section, we present our numerical estimates of the thresholds of $\bbZ_d$-KTCs for $2\leq d\leq6$ subject to the generalized bit-flip noise model introduced in the previous section. The threshold is defined as the value of the physical noise rate $p_{\rm phys}$ below which the decoding error probability $p_{\rm dec}$ can be made arbitrarily small by increasing the lattice size $L$.

The simulations were performed as follows.  For various values of $d$, $L$ and $p_{\rm phys}$, specifiying a $\bbZ_d$-KTC of linear size $L$ subject to a noise of parameter $p_{\rm phys}$, we performed a Monte Carlo simulation to estimate the decoding error probability $p_{\rm dec}$. We used sample sizes of the order of $10^4$. For a fixed value of $d$, we plotted estimates of $p_{\rm dec}$ vs $p_{\rm phys}$ for different values of $L$. We then used the fitting model $p_{\rm dec}=(p_{\rm phys}-p_{\rm th})L^{1/\nu}$ (see \cite{DKLP02a,H04a} for more details) to estimate the value of the threshold. As an example, we plotted the results and the fits for $\bbZ_3$-KTC on \fig{Z3_pth}.  

Repeating this for $3\leq d\leq6$ (2 was studied in \cite{DP09a,DP10a}), \fig{HB_Results} shows $p_{\rm th}$ as a function of $d$. Heuristically, we did expect that the value of $p_{\rm th}$ increases with $d$. Indeed, if we imagine simulating a qudit using $\log_2 d$ qubits, a fixed noise rate for increasing values of $d$ translates into a decreased noise rate per qubit. Moreover, it was reported in \cite{AMT12a} that the performance of BP for $\bbZ_d$-KTC, which is very poor in the qubit case, is greatly increased as $d$ grows.

It is intringuing to note that for $\bbZ_2$-KTC subject to bit-flip or depolarizing noise, $p_{\rm th}$ is numerically very close to the hashing bound \cite{DKLP02a,H04a,BAOKM12a}. The hashing bound, obtained by a simple packing argument \cite{EM96a}, states that for non-degenerate CSS codes,
\begin{align}
	0 & \leq 1-2H_2(p), \label{eq:HB2}
\end{align}
where $H_2$ is the binary entropy: $H_2(p)=(1-p)\log_2(1-p)+p\log_2 p$. From \eq{HB2}, one can calculate the saturating point $C_2\approx0.110$ which is indeed quite close to the optimal threshold of the $\bbZ_2$-KTC subject to independent bit-flip and phase-flip errors, $p_{th}(2)\approx0.109(4)$ \cite{DKLP02a,H04a}. This near coincidence is intriguing given that topological codes are highly degenerate, so there is no reason they should obey the hashing bound. Of course, the decoder we are using here is sub-optimal, so the threshold we find $p_{th}(2)\approx0.89(6)$ is a smaller fraction $\alpha = p_{\rm th}(2)/C_2 \approx 0.81(4)$ of the hashing bound.

For qudits, the hashing bound is
\begin{align}
	0 & \leq 1-2H_d(p) \label{eq:HBd},\\
	\textrm{with}\quad H_d(p)&=(1-p)\log(1-p)+p\log{\frac{p}{d-1}}. \nonumber
\end{align}
In this case, we find $C_3\approx0.159,C_4\approx0.189$ and so on.  Figure~\ref{fig:HB_Results} shows the threshold $p_{\rm th}(d)$ obtained with the RG decoder as well as a rescaled hashing bound $\alpha C_d$ where $\alpha$ is determined by the $\bbZ_2$ fit. The agreement is both unexplained and  surprisingly good. Note also that even though our decoder is sub-optimal, $p_{th}(d+1)>C_d$ for all $d$ we have studied, which strongly support the claim that the threshold increases with $d$.

\section{Conclusion}
In this paper, we presented a generalization of the renormalization group decoder of \cite{DP09a,DP10a} to Kitaev topological codes built with the groups $\bbZ_d$. Our numerical results show that the threshold value increases as a function of the local dimension $d$. Moreover, its behaviour is in very good agreement with a scaling predicted by the hashing bound. This trend could be confirmed by more accurate numerical estimates using a mapping to a statistical mechanics model, which does not require solving the decoding problem \cite{DKLP02a, BAOKM12a}. A theoretical understanding of this behavior is also desirable. Lastly, estimating the threshold in the presence of measurement error and detailed syndrome measurement circuits on qudits remains an interesting open question.

\section{Acknowledgements}
We would like to thank Jonas Anderson for useful discussions regarding the generalized hashing bound. We also thank Simon Burton, Courtney Brell and Stephen Bartlett for enlightening discussions of Kitaev's construction \cite{K03a}. Computational resources were provided by Calcul Qu\'ebec and Compute Canada. This work was partially funded by NSERC and by Intelligence Advanced Research Projects Activity (IARPA) via Department of Interior National Business Center contract D11PC20167. The U.S. Government is authorized to reproduce and distribute reprints for Governmental purposes notwithstanding any copyright annotation thereon. Disclaimer: The views and conclusions contained herein are those of the authors and should not be interpreted as necessarily representing the official policies or endorsements, either expressed or implied, of IARPA, DoI/NBC, or the U.S. Government.

\bibliographystyle{apsrev}
\bibliography{Zn.bib}

\end{document}